\documentclass[%
superscriptaddress,
amsmath,amssymb,
aps,
notitlepage
]{revtex4-1}
\usepackage[hidelinks=true]{hyperref} 
\hypersetup{
  colorlinks   = true, 
  urlcolor     = blue, 
  linkcolor    = blue, 
  citecolor   = red 
}
\usepackage{tikz,pgfplots}
\usepackage{textcomp}
\usepackage{gensymb} 
\usetikzlibrary{external}
\pgfplotsset{
  compat=newest,
  invoke before crossref tikzpicture={\tikzexternaldisable},
  invoke after crossref tikzpicture={\tikzexternalenable},
}
\usetikzlibrary{
    external,
}
\pgfplotsset{plot coordinates/math parser=false} 
\usetikzlibrary{plotmarks}
\usepackage{caption}
\usepackage{siunitx}
\usepackage{subcaption}
\usepackage{graphicx}
\usepackage{dcolumn}
\usepackage{bm}

\def\sgp{\color{\red}}

\def\t{\mathbf{t}}
\def\n{\mathbf{n}}
\def\r{\bm{r}}

\def\x{\bm{e_x}}

\def\k{\kappa}
\def\ee{\mathcal{E}_{el}}
\def\nee{\bar{\mathcal{E}}_{el}}
\def\hk{\hat{\kappa}}

\def\sgp{\textcolor{black}}

\begin{document}


\title{Relaxation of a highly deformed elastic filament at a fluid interface}
\author{S Ganga Prasath}
\email{sgprasath@tifrh.res.in}
\affiliation{TIFR Centre for Interdisciplinary Sciences, Tata Institute of Fundamental Research, 21 Brundavan Colony, Narsingi, Hyderabad 500075 India.}
\author{Joel Marthelot}
\affiliation{TIFR Centre for Interdisciplinary Sciences, Tata Institute of Fundamental Research, 21 Brundavan Colony, Narsingi, Hyderabad 500075 India.}
\author{Rama Govindarajan}
\affiliation{TIFR Centre for Interdisciplinary Sciences, Tata Institute of Fundamental Research, 21 Brundavan Colony, Narsingi, Hyderabad 500075 India.}
\author{Narayanan Menon}
\affiliation{TIFR Centre for Interdisciplinary Sciences, Tata Institute of Fundamental Research, 21 Brundavan Colony, Narsingi, Hyderabad 500075 India.}
\affiliation{Department of Physics, University of Massachusetts, Amherst, MA 01003-3720 USA.}

\date{}
\begin{abstract}
We perform experiments to investigate the relaxation of a highly deformed elastic filament at a liquid-air interface. The dynamics for filaments of differing length, diameter and elastic modulus collapse to a single curve when the time-dependence is scaled by a time scale $\tau = 8 \pi \mu L_o^4/B$. \sgp{Even though the time $\tau$ is obtained by comparing the linear bending and viscous forces,  we find that it also controls the relaxation in the highly nonlinear regime of our experiments. The relaxation, however, is completed in a very small fraction of the time $\tau$ due to a prefactor that changes with the tension in the nonlinear regime.} Nonlinear numerical simulations show that the force due to tension along the filament is comparable to the bending force, producing a net elastic restoring force that is much smaller than either term. We perform particle image velocimetry at the liquid-air interface to support the results of the numerics. Finally, we find that when the filament is initialized in asymmetric shapes, it rapidly goes to a shape with symmetric stresses. This symmetrisation process is entirely non-linear; we show that the symmetric curvature state minimizes energy at arbitrarily large deformation.
\end{abstract}

\pacs{Valid PACS appear here}
\maketitle
\section{\label{sec:intro} Introduction}
Highly deformed slender elastic filaments are to be found across several decades of length scales starting from crops and tree canopies in wind \cite{de2008effects}, aquatic plant stems in flowing streams \cite{miler2012biomechanical}, propelling flagellae of organisms \cite{purcell1977life,lighthill1976flagellar,hyams1978isolated,yang2009kinematics,2008PhFl...20e1703C}, stereocilia inside ears \cite{cotton2004computational,kozlov2011forces}, to suspensions of fibres~\cite{quennouz2015transport,evans2013elastocapillary,manikantan2014instability} (see ref.~\cite{lindner2014elastic} for a recent review). The most heavily studied of these examples is the driven dynamics of flagella \cite{wiggins1998flexive,2006PhFl...18i1701Y}, where the balance is between forces due to bending, which tend to straighten the filament, and viscous drag, which acts to damp the motion.

Unless they are held in that state by external or internal forces, filaments will relax from a highly bent state to their equilibrium, stress-free state. For a filament with bending modulus $B$ and length $L_o$, the bending energy \sgp{per unit length} is $B \kappa^2$, quadratic in the local curvature $\kappa$. This leads to bending moments and forces \sgp{per unit length} $\sim BL_o^{-3}$ that are linear in the displacement from the unstressed conformation of the filament \cite{landau1959course}. Balanced against the drag force \sgp{per unit length} $8\pi\mu L_o/\tau$ from a fluid of viscosity $\mu$, we obtain a characteristic time scale for dynamics over the length of the filament: $\tau = 8 \pi \mu L_o^4/B$. However, when the deformations are large, another source of  stress becomes significant:  in order to satisfy the constraint of constant length,  a gradient of tension appears along the filament. This tension is a nonlinear function of the geometry \cite{goldstein1995nonlinear}.  Previous work has concentrated on bending alone, and the role of this nonlinearity is largely unexplored. The goal of this work is to understand the relaxation of a highly-deformed filament from its high elastic-energy state, with full consideration of the nonlinear effects of the geometry.

We perform experiments to study the relaxation to a straight configuration of an initially highly deformed elastic filament. The dynamics are restricted to a two-dimensional plane by placing the filament at the interface of a viscous fluid. The filament is initially deformed by holding its ends in place with two fine needles at the interface. When these are removed, it relaxes towards a straight, unbent shape.  Fig.~\ref{fig:expt_setup}$(a)$ shows the initial deformed state and several intermediate steps in this relaxation process.  We vary the parameters of this system -- the length, diameter and material of the filament -- to understand the time-dependence of this process. In order to probe internal variables such as the tension, we solve a numerical model  \cite{goldstein1995nonlinear,evans2013elastocapillary} to compute elastic forces in the filament. This model includes a fully nonlinear treatment of the elasticity of the filament, but a simplified description of the hydrodynamic drag. We also perform Particle Image Velocimetry to visualise the flow-field around the filament and validate the results of the theoretical model.

\section{Experimental methods}
\sgp{The filaments we use are made of a elastomer, vinyl polysiloxane (VPS), which was prepared with two different Young's moduli, $E$ (of $240 kPa$ and $ 800 kPa$) whose densities are $1020 kg/m^3$, $1180 kg/m^3$}. The precursor material is injected into a capillary tube which defines the diameter $d$ of the filament. Once the polymer cures, the filament is extracted from the tube and cut to the desired length $L_o$.  All deformations of the elastomer are fully reversible.

\begin{figure*}
\centering
\includegraphics[scale=0.8]{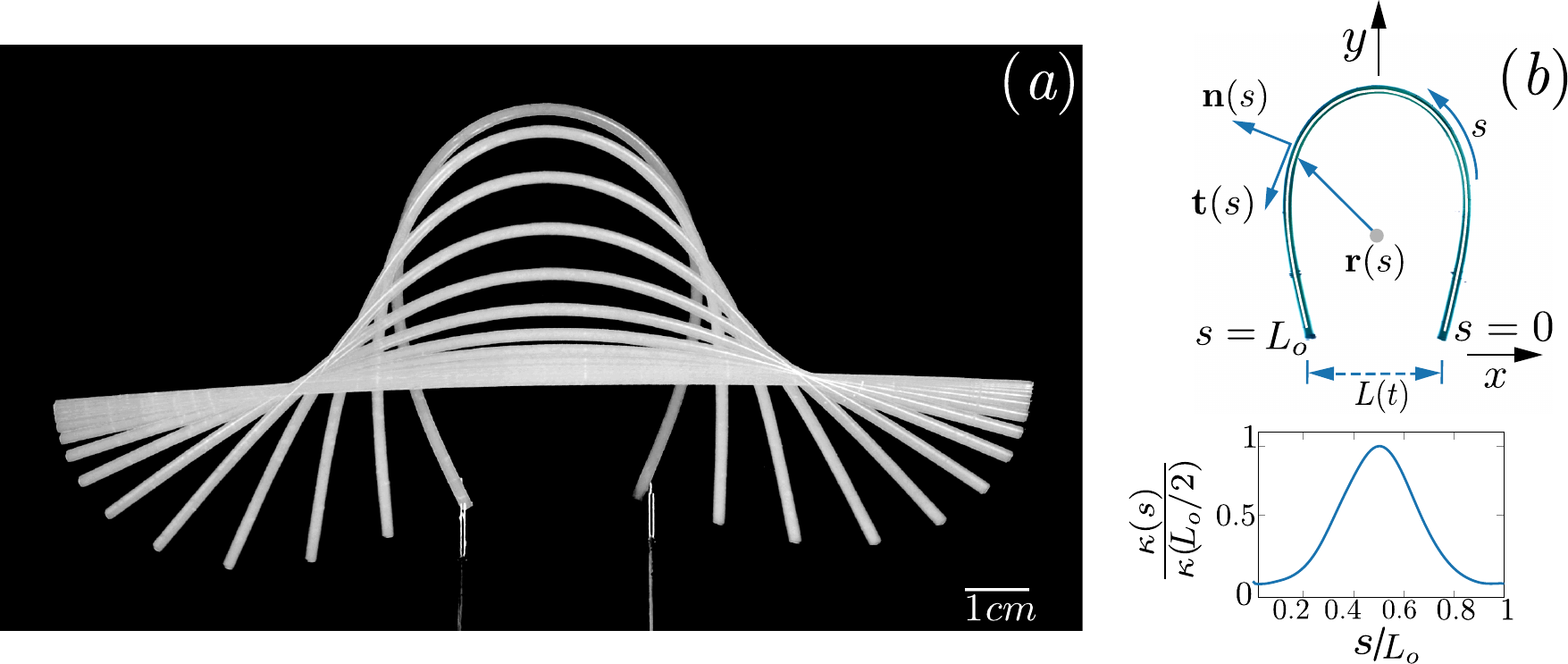}
\caption{$(a)$ Superimposed images of the filament taken at intervals of 1 $sec$. The two needles at the bottom of the image are used to release  the filament from its initial configuration. $(b)$ The thick white line over the filament (seen as the cyan outline) is the Bezier fit. We plot below this the corresponding curvature computed with this fit. Here the position vector is $\r(s)$ and the unit vectors $\n(s)$ and $\t(s)$ in the normal and tangential directions.}
\label{fig:expt_setup}
\end{figure*}

This filament is placed on the surface of a cylindrical dish of glycerol. The ends of the filament are held, then released by needles attached to a tweezer mechanism mounted on a translational stage. \sgp{The density of glycerol = $1216 kg/m^3$, and the air-glycerol surface tension = $64mN/m$; thus the interface is not significantly deformed by gravity. The depth of immersion of the rods is determined by the contact angles with glycerol, which were determined to be $102.4\degree$ and $94.6\degree$ respectively for the stiff and the soft VPS. Thus approximately half the filament's surface area is submerged. The filament stays on the interface and the dynamics are fully 2-dimensional. No twist occurs in the experiments.} 
 The high viscosity ($\mu = 1.412 Pa.s$ at $100\%$ concentration) keeps the dynamics in the stokesian regime; \sgp{the Reynolds number $Re =\rho u d/\mu \approx 10^{-2}-10^{-3}$} immediately after release, when the filament is moving at its fastest. As shown in fig.~\ref{fig:expt_setup}$(a)$, at no time in the relaxation process does any part of the filament come close to self-contact (or to the walls of the dish), so capillary forces can be neglected. 

We track the filament shape as a function of time using a Nikon D5000 camera at a resolution of $4288 px \times 2848 px$ and a frame rate of $1 fps$. As shown in fig.~\ref{fig:expt_setup}$(b)$, from the images we extract $\r(s,t)$, the position vector of the filament centerline along the arc length $s$, using the following procedure. We first separate the filament from the background, and then reduce these pixels to a set of equally spaced points.

In order to accurately take higher order derivatives with respect to material coordinates, we make a Bezier fit to these points. This is a unique polynomial fit of $\mathcal{O}(l^{n})$ to a given set of $n$ points in $\r(s)=\mathcal{B}\{x(l),y(l)\}$, $l$ being the parameterisation of the Bezier curve. As $l$ goes from 0 to 1, $s$ goes from 0 to $L_o$, but note that the two are not linearly related. The analytical form of the fit allows us to calculate $\kappa(s)$  (fig.~\ref{fig:expt_setup}$(c)$) as a continuous function. The curvature is given by
$$
\kappa(s) = \frac{x'y''-y'x''}{(x'^2+y'^2)^{3/2}},
$$
where the primes denote differentiation with respect to $l$. We also compute the elastic energy $\ee(t)$ [$=(1/2)\int_0^{L_o} B \k^2(s) ds$] from the curvature profiles extracted from the images. 

\section{Experimental results}
\begin{figure*}
\centering
\includegraphics{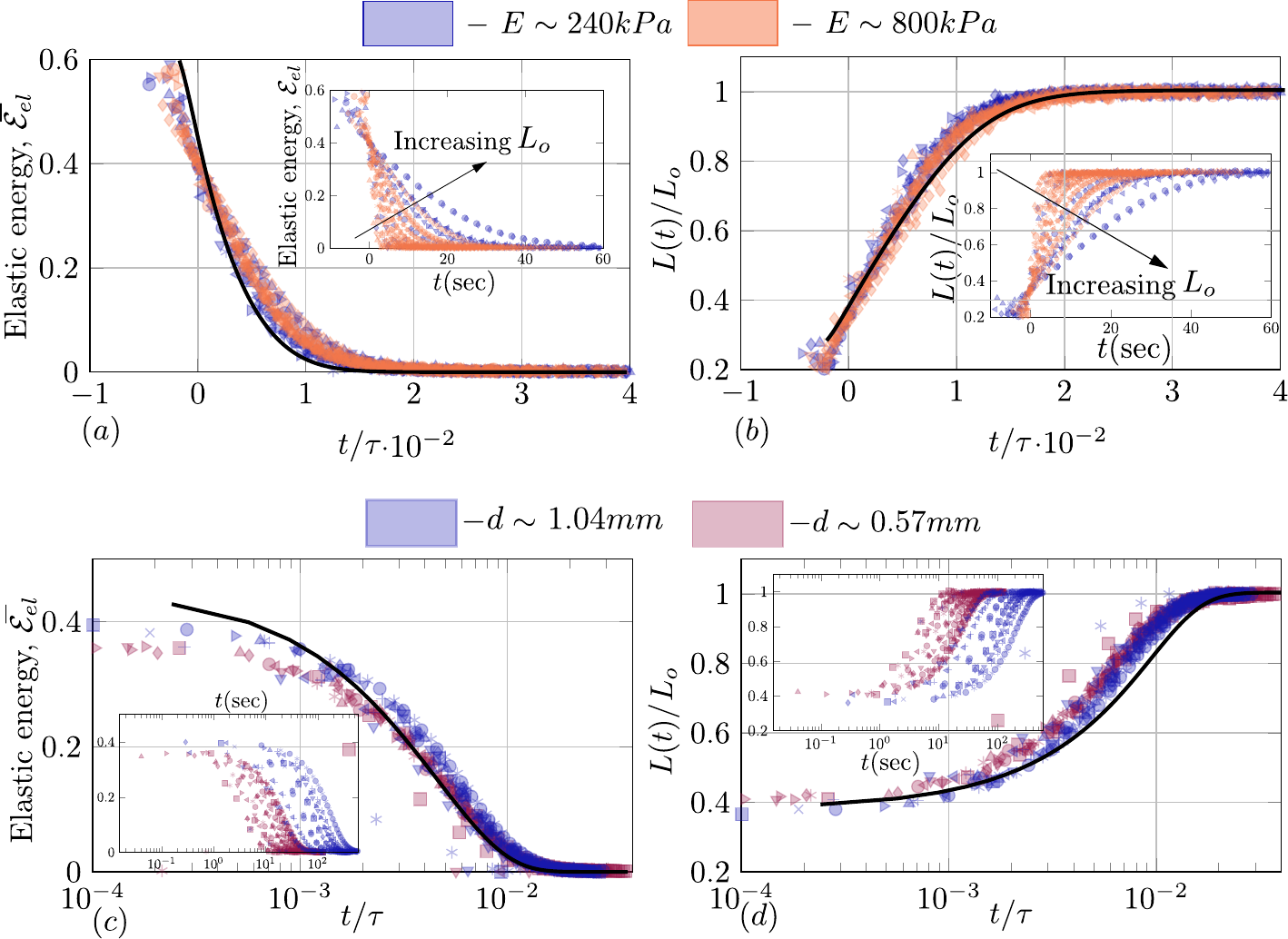}
\caption{$(a)$ Normalised elastic energy, $\nee$ and $(b)$ non-dimensional end-to-end distance, $L(t)/L_o$ for two different values of Young's modulus, $E$ ($240 kPa \ {\rm and} \ 800 kPa$) and different lengths ($L_o$ varied from $4.1cm$ to $7.3cm$ for 240kPa and from $4.6cm$ to $7.8cm$ for 800kPa). The insets show the data as functions of dimensional time, whereas the main figures are plotted in terms of scaled time $t/\tau$. $(c), (d)$ Similar plots to that of $(a), (b)$ for two different diameters, $d - 0.57 mm, 1.04 mm$ ($E = 240 kPa$) and $L_o$ varied between $4.6cm-7.2cm$ for the former and $3.3cm-6.3cm$ for the latter. The solid lines in all the subfigures indicate results obtained by solving eqs.~\ref{eq:position_dl}, \ref{eq:tension_dl} numerically, and scaled by a factor of 4.2 in time, as discussed later in the text.}
\label{fig:length_young}
\end{figure*}

Since the elasticity of the filament is determined by its bending modulus $B=E\pi d^4/4$, we vary $d$ and $E$  to study the dependence of the relaxation time on these filament parameters. We also vary Lo;  since the distance between the needles holding the filament in its initial configuration is fixed the initial average curvature, and the relative initial separation between the ends, both vary. From the sequence of images that characterize the shapes of the relaxing filament, we extract two relevant physical quantities. One observable is the non-dimensional end-to-end distance, $L(t)/L_{o}$ (see fig.~\ref{fig:expt_setup}), as a function of time. We also report the elastic energy $\nee(t)$, normalized by that of a filament of length $L_o$ rolled  into a circle$: 2B\pi^2/L_o$. These quantities are plotted in the insets of fig.~\ref{fig:length_young}$(a, b)$), for two different values of Young's modulus, and for several values of the length $L_o$. It is apparent that the relaxation time increases monotonically and strongly with $L_o$ and decreases with increasing $E$. 

As shown in the main fig.~\ref{fig:length_young}$(a, b)$, all the data for length and elastic energy collapse on a single scaled curve. The time has been scaled in each case by $\tau = 8\pi \mu B^{-1}L_o^4$. Even though we are in a very nonlinear regime, these curves collapse when plotted in terms of $t/\tau$. We remark that the time scale $\tau$ has been obtained merely by balancing viscous and bending forces in the linear regime. In all the data in fig.~\ref{fig:length_young} we have chosen the origin of time $t=0$ as the instant where non-dimensional elastic energy, $\nee = 0.4$ but the data collapse even at negative times thus defined. Moreover, upon choosing just $\nee(t=0)$ as $0.4$, the respective end-to-end distances automatically collapse at the initial time. The value of 0.4 is arbitrary and the scaling works well for any other origin of time. We note that most of the relaxation is accomplished when the nondimensional time is very small, that is, $t/\tau \sim 2\times 10^{-2}$ rather than $t/\tau \sim \mathcal{O}(1)$. 

We also vary the bending modulus through its strong dependence on $d$, the diameter of the filament.  We choose two different diameters, $d = 1.04 mm$ and $0.57 mm$ with Young's modulus fixed at $E = 240 kPa$, at several different lengths.  The data in the insets of fig.~\ref{fig:length_young}$(c,d)$ track the measured relaxation dynamics and show that the  smaller diameter relaxes slower.
Once again, the relaxation dynamics collapses when data obtained by varying $L_o$ and $d$ are plotted against the scaled time, $t/\tau$ as shown in fig.~\ref{fig:length_young}$(c,d)$. The slowest and the fastest dynamics span a factor of $27$ variation in time scale $\tau$. 

There are two major findings in our experiments. First, the time scale $\tau$ used to collapse the data 
is taken from the balance of forces in the linear regime, even though most of the dynamics we probe are in the deeply nonlinear regime, where curvature $\sim L_o^{-1}$. Second, the time-scale of relaxation is about two orders of magnitude smaller than $\tau$. A deeper understanding of this puzzle requires us to quantify the dynamics in the nonlinear regime. The nonlinear regime is governed by the internal tension, which is not accessible in our experiments. We thus turn to a numerical simulation of the fully nonlinear  equations discussed below.  

\section{Nonlinear dynamical equations}

The dynamical equation for an elastic filament (see \cite{goldstein1995nonlinear,2004JCoPh.196....8T} for more details) can be derived by a variational formulation. The mechanical energy of bending as well as a constraint term to enforce length conservation yields an elastic energy of:
\begin{equation}
\ee = \frac{B}{2} \int_0^{L_o} \kappa^2(s) ds + \int_0^{L_o} \frac{T(s)}{2} \{ |{\t}(s)|^2 -1 \} ds
\end{equation}
where  ${\t}(s)\equiv\r_s$ is the tangent vector and the tension, $T(s)$, enters as a Lagrange multiplier ensuring inextensibility. The elastic force inside the filament is obtained from the corresponding Euler-Lagrange equations:
\begin{equation}
\bm{f}_{el} = -B \r_{ssss} + \partial_s[T(s) \r_s].
\end{equation}
The boundary conditions are: $\r_{ss}(0)=\r_{ss}(L_o)=0$ corresponding to zero moment and $\r_{sss}(0)=\r_{sss}(L_o)=0$, $T(0)=T(L_o)=0$, corresponding to zero force at the free ends.
In an over-damped situation, the viscous force balances the elastic force:
\begin{equation}
8\pi\mu\partial_t \r = -B \r_{ssss} + \partial_s[T(s)\r_s].
\label{eq:dynamics}
\end{equation}
After non-dimensionalising $\r$ by $L_o$, and $t$ by an arbitrary time scale $\tau_1$, the equation of motion is:
\begin{equation}
\bar{\mu}\partial_t \r = - \underbrace{\r_{ssss}}_{\mathcal{F}1} + \underbrace{\partial_s[T(s)\r_s]}_{\mathcal{F}2}.
\label{eq:position_dl}
\end{equation}
The non-dimensional parameter $\bar{\mu} = 8\pi\mu {L_o}^4/(B \tau_1)$ is the ratio of viscous force to bending force. Note that when we set $\tau_1=\tau$, we get $\bar\mu=1$, and the above equation becomes parameter-free.
Using the inextensibility constraints, $\partial_s|\t^2|=0$ and $\partial_t|\t^2|=0$, the equation for tension reads as:
\begin{equation}
(\partial_{ss} - |\r_{ss}|^2)T(s) = -(3 |\r_{sss}|^2 + 4(\r_{ss}\cdot\r_{ssss})).
\label{eq:tension_dl}
\end{equation}
Equations (4) and (5) constitute the basic equations for the mechanics of the filament. There are two approximations in these equations for the dynamics of the filament centreline. First, the hydrodynamic interaction between points on the filament is neglected. Second, the drag force due to the motion of the filament is assumed to be isotropic. \sgp{We have checked computationally that the anisotropic drag, as used in Quennouz et al.~\cite{quennouz2015transport}, has negligible effect on the dynamics. This is because the motion of the filament is predominantly in the direction normal to it. We therefore proceed by using an isotropic drag coefficient, which does not depend on the length and diameter of the filament.} Furthermore, in our experiments, we do not have an independent measure of the drag at the interface (which will depend on wetting properties among other things), and so in the theoretical computations of drag, we use the value of the bulk viscosity of the liquid. An effective viscosity at the interface contains an unknown factor to be determined experimentally.

\section{Nonlinear simulation}

We solve numerically these fourth-order, nonlinear equations  with our simplified model for the hydrodynamic drag. 

\subsection{Numerical method and validation}
\begin{figure*}
\centering
\includegraphics[trim={40 0 0 0},scale=0.85]{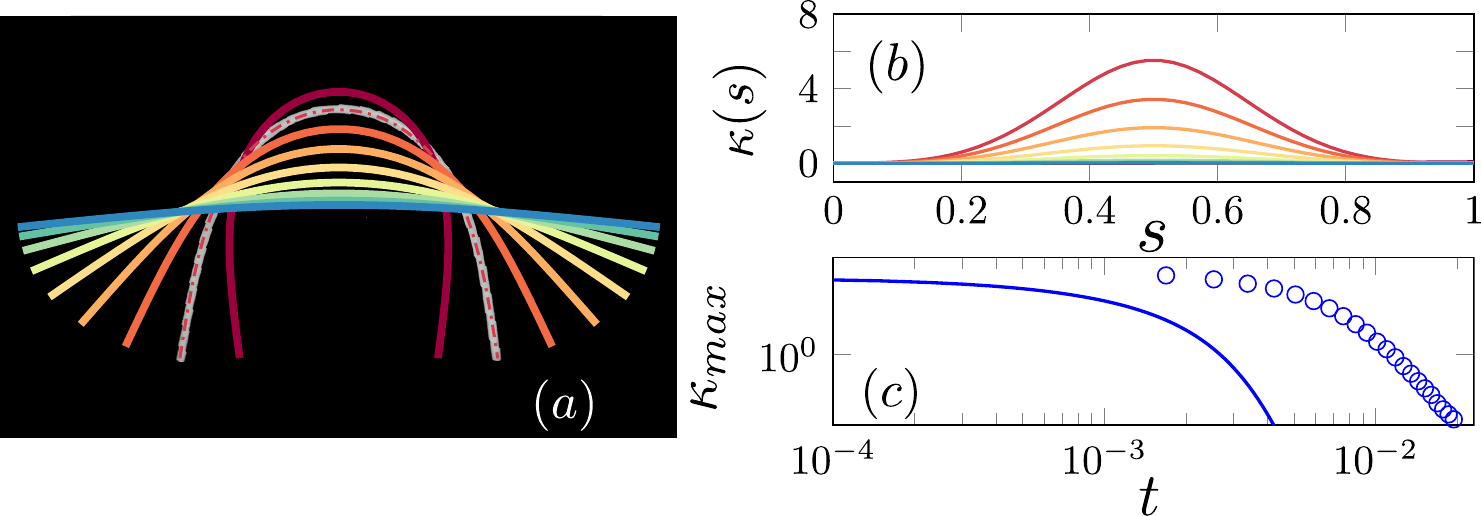}
\caption{$(a)$ Shape of the filament for various time instants in the simulation super-imposed on each other, with an experimental image, seen in white, in the background. $(b)$ Evolution of curvature, $\k(s)$ vs $s$ for different times. $\k(s)$ is always positive and decays monotonically with time. $(c)$ The decay of the maximum value of curvature compared with that of experiments ($\circ$) without any fitting parameter in the numerics. All quantities are non-dimensionalised using $L_o$ and $\tau$ as scales. 
}
\label{fig:time_evolution}
\end{figure*}

As in \cite{2004JCoPh.196....8T} we solve  the tension equation as well as the equation of motion by discretizing the filament into inter-connected rods of length $ds$ while conserving the total length using penalisation. We use a TDMA scheme to calculate $T(s_i)$ using the value of $\r(s_i)$ at discrete arc-lengths, $s_i$. A skew-finite difference is used for implementing the boundary conditions in higher-order derivatives. 

Fig.~\ref{fig:time_evolution}$(a)$ shows the evolution of the computed filament shape, compared to an image from the
experiment. More quantitatively, we go back to fig.~\ref{fig:length_young}$(a,b)$, where we show $\nee(t)$ vs ${t/\tau}$ and $\bar{L}(t)$ vs ${t/\tau}$ computed by setting $\bar{\mu}=1$.
 The time-evolution of the shape agrees well with the experimental data. The evolution of $\bar{L}(t)$ (see eq.~\ref{eq:lt} in the appendix) depends only on the gradient of $T(s)$ at the boundary. The fast evolution of $\bar{L}(t)$ at early times is a consequence of $T(s)$ having a gradient of large magnitude at the boundary. However, the simulation result is scaled from the experimental data by a factor of $~4.2$. \sgp{As discussed earlier, the filament is only half submerged in the fluid. A factor of about $2$ arises from this, and explains in part the difference between simulations and experiment. The remaining discrepancy could be due to (i) neglect of hydrodynamic interaction (ii) neglect of capillary effects, and potentially most importantly (iii) our use of the bulk viscosity of the liquid in the model even though the drag occurs at an interface. The effective viscosity at the interface is expected to be less than in the bulk, as glycerol is hygroscopic. We emphasize that the idealizations in the numerical model are in the treatment of drag, and not in the elasticity of the filament.}

Indeed, fig.~\ref{fig:time_evolution}$(b)$ shows that the computed curvature is always positive along the filament. The maximum of the curvature as a function of time is shown in fig.~\ref{fig:time_evolution}$(c)$, where it is compared to the experimental data. As discussed earlier, the time evolution of
experiment and numerical computation have the same functional form, but the time scale
observed in the numerics is faster. 

\subsection{Tension}
\begin{figure*}
\centering
\includegraphics[trim={20 0 0 0},scale=1.0]{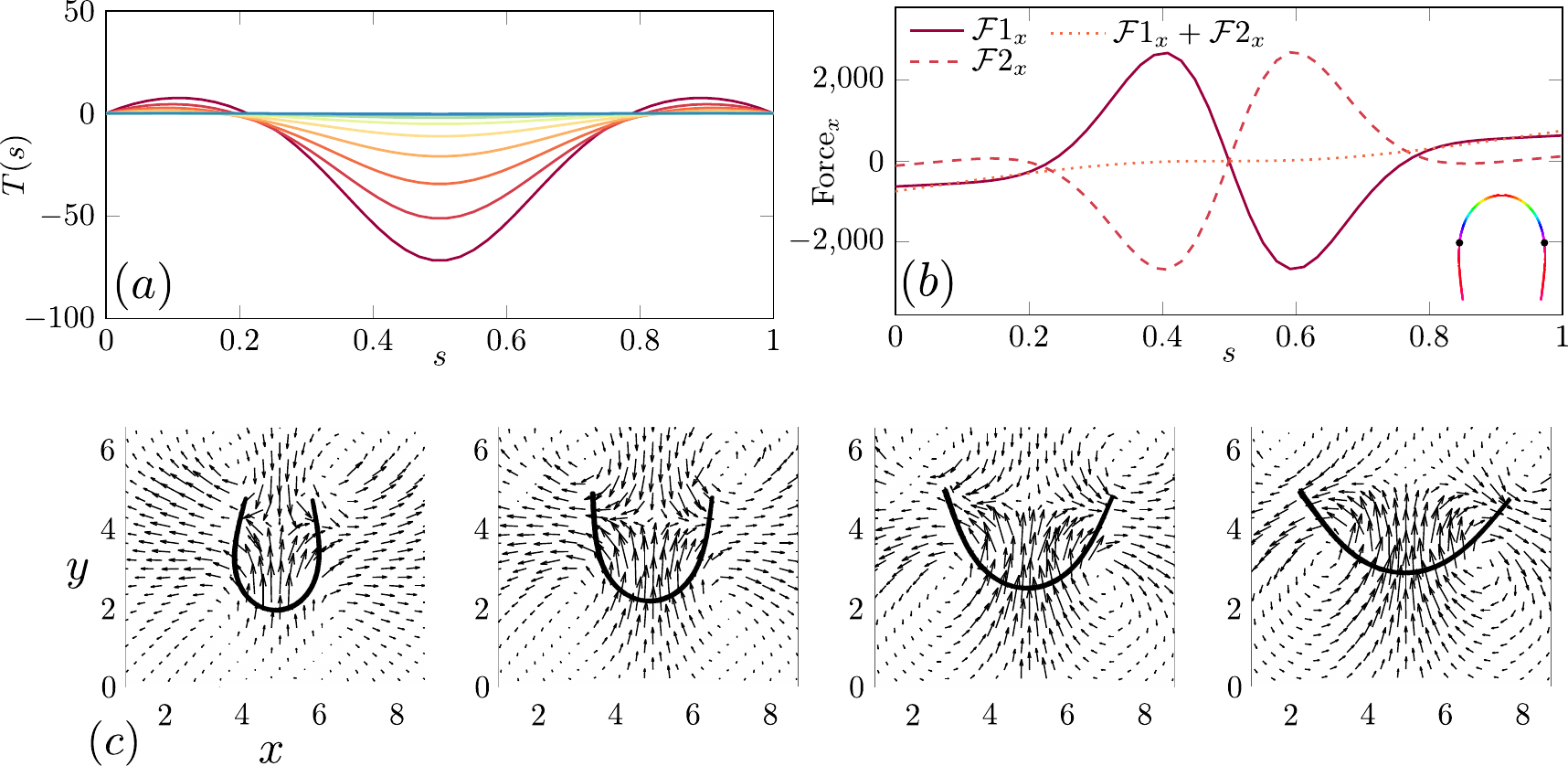}
\caption{$(a)$ Tension, $T(s)$ vs $s$ at various time instants. The profile of tension changes shape as it decays, which is evident from the fact that the position where it crosses zero travels towards the ends in time. $(b)$ $x$-component of force due to bending ($\mathcal{F}1$), tension ($\mathcal{F}2$) and the resultant viscous force (their sum) from eq.~\ref{eq:dynamics} for the filament configuration shown at the bottom right. The contribution from bending force and that of tension are very similar in magnitude but opposite in sign, so a small viscous force (as seen) is sufficient to balance them. The filament in the inset is colored based on $T(s)$ and the dots show the position where $T(s)$ goes to zero. $(c)$~Velocity field obtained by PIV where a pair of vortices is formed on each side of the filament. The tension in the filament vanishes near the location where the sign of vorticity changes. Scales in units of $cm$.}
\label{fig:forces}
\end{figure*}

Having validated our model, we now move to computed quantities that are not accessible in the experiments. As is clear in figure~\ref{fig:time_evolution}, the curvature is not uniform, and therefore the filament should not be expected
to open uniformly. In fig.~\ref{fig:forces}$(a)$ we show the tension, $T(s)$ computed using eq. \ref{eq:tension_dl}. The relationship between curvature and tension is not linear and depends on higher derivatives of curvature. Consequently, unlike the curvature, which is always positive, the tension changes sign from  compressive in the central region to tensile near the free ends. Force balance along the tangential direction thus requires that the viscous drag force be a function of arc-length.

In order to test whether the tension distribution observed in the simulation results are supported by experimental observations of the viscous drag on the filament, we use Particle Image Velocimetry (PIV) to measure the velocity field at the interface. We make a suspension of hollow glass spheres of $10 \mu m$ average diameter in the fluid and illuminate the interface with laser sheets (\textsc{Wicked Lasers} $<500mW$ and wavelength $\approx 532nm$). The motion of the particles from the recorded images is tracked using open-source package, \textsc{PIVLab} \cite{thielicke2014pivlab}. The resultant velocity field for the relaxation is plotted in fig.~\ref{fig:forces}$(c)$. As the filament begins to relax, two symmetric pairs of vortices are formed. Tension in the system has contributions only from the tangential projection of the stress tensor. Now using eq.~\ref{eq:position_dl} we can write:
\begin{align*}
\r_s \cdot \hat{\t} \ \partial_s T(s) &= \bar{\mu} \partial_t \r \cdot \hat{\t} + \r_{ssss} \cdot \hat{\t} \\
T(s) &= \int_0^s (\r_{ssss}\cdot \hat{\t} + \bar{\mu}\r_t \cdot \hat{\t}) ds
\end{align*}
This quantity goes to zero at the position where net bending force is balanced by viscous dissipation. Comparing the PIV field and  fig.~\ref{fig:forces}$(b)$ we observe that this location is close to the point where vorticity changes sign. This location moves towards the ends of the filament as it relaxes, consistent with the simulation results. 
  
\sgp{To separate the contributions of the bending force, $\mathcal{F}1$, and the tension arising from length conservation  $\mathcal{F}2$, we plot in fig.~\ref{fig:forces}$(b)$ $x$-component of these individual terms on the right side of eq.~\ref{eq:dynamics}. The bending force ($\mathcal{F}1$) and force due to tension ($\mathcal{F}2$) are comparable in magnitude, though opposed in sign, even for the highly deformed regime.} The algebraic sum of these forces is balanced by a viscous term that is much smaller in magnitude than each of these terms.

Viewed in the light of these results, it is evident that ignoring the tension term in the large deformation limit not only leads to non-conservation of length but leads to a significant overestimation of the dissipation due to viscosity. 
Second, as we see from the energy functional, $T \sim B/L_o^2$, the nonlinear tension term ($\mathcal{F}2$)  has the same scaling as the bending term ($\mathcal{F}1$). \sgp{In other words, $\tau$ is the single time scale in the problem, even in the nonlinear regime but with a small coefficient due to tension.} We recall that the filament relaxes much faster experimentally than the time scale $\tau$, i.e., the balance between the two terms leads to a small numerical prefactor. 

\section{Asymmetric initial conditions}

\begin{figure*}
\centering
\includegraphics[trim={40 0 0 0}]{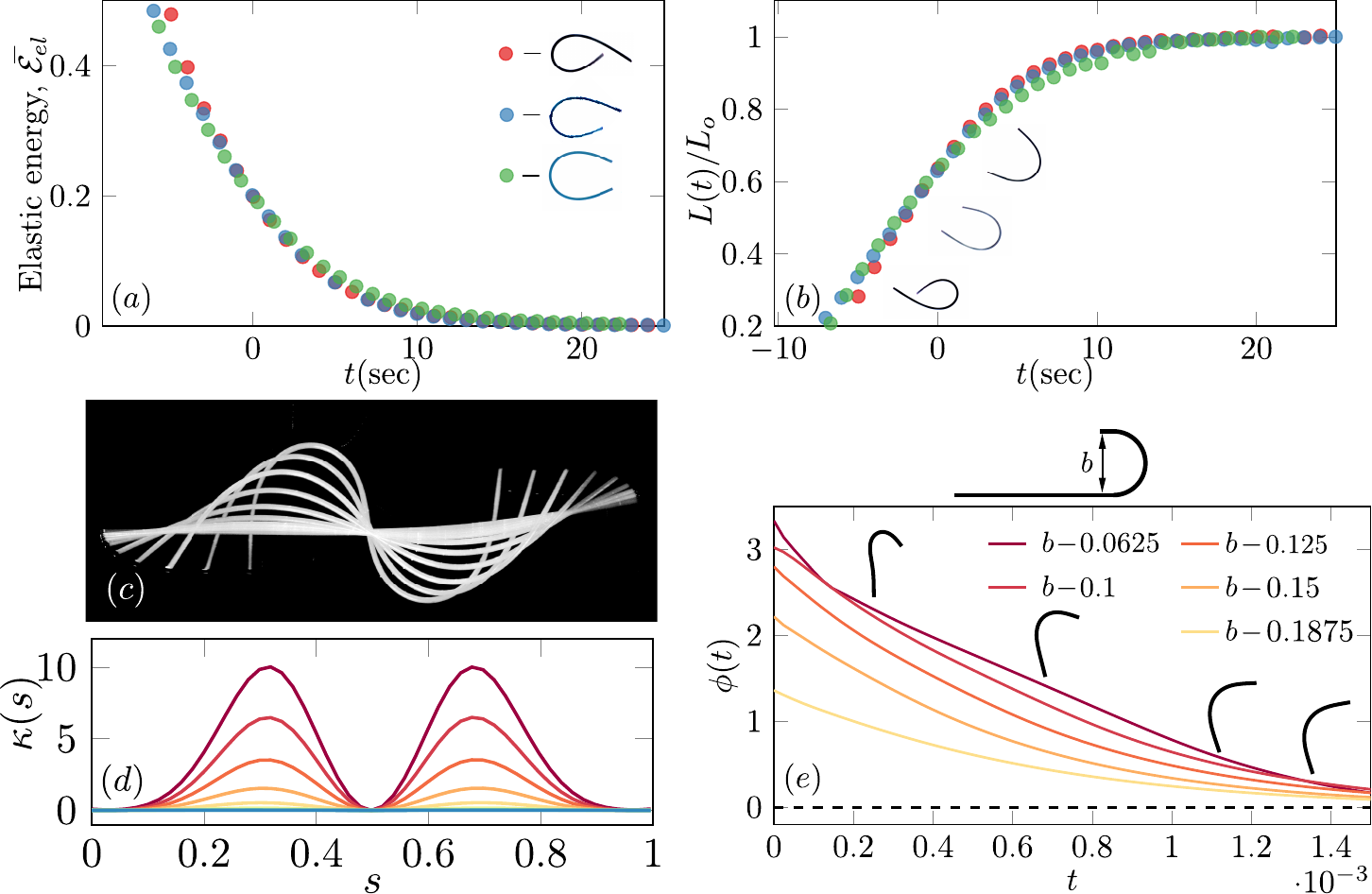}
\caption{$(a)$ Evolution of elastic energy and $(b)$ $L(t)/L_o$ for the two asymmetric initial conditions shown in the legend from experiments, compared with the symmetric case ($L_o=6.5cm, E=240kPa, d=1.1mm$). We see that the curves merge quickly and the decay beyond that time becomes identical. $(c,d)$ Evolution of a rod with symmetric $\k(s)$ about $s=0.5$ which relaxes to a straight configuration along a symmetric path. $(c)$ consists of superimposed images from experiment while $(d)$ is from numerics for the same initial condition. $(e)$ Non-dimensional parameter $\phi(t)$ which quantifies the asymmetry in a given configuration. This is plotted for different initial conditions with different values of $b$, where $b$ is the diameter of the semi-circular section shown above the plot. $b$ here is varied between 0.0625 and 0.1875. We see that a configuration with smaller $b$ symmetrises faster.}
\label{fig:asymm_data}
\end{figure*}

All the results we describe above are for initial shapes that are symmetric about the midpoint of the filament ($\k(L_o/2-s) = \k(s)$). If the filament is released from an asymmetric, highly deformed shape, we find experimentally that it first deforms into a symmetric shape, then relaxes along the sequences of symmetric shapes we have previously shown.
This is shown in fig.~\ref{fig:asymm_data}$(a, b)$ via the evolution of $\nee(t)$ and $\bar{L}(t)$. The images of filaments shown label the initial condition in each of three data sets displayed. We have chosen $t=0$ to be the point where there is no experimentally discernible asymmetry. For $t > 0$ the relaxation follows the same path in all cases. 

In the linear regime, \sgp{contributions from tension becomes negligible} and the relaxation time of a Fourier component varies only with wavelength, and any initial asymmetry would be preserved. Thus we emphasize here that the collapse into a symmetric shape is a consequence of being in a deeply nonlinear regime. The linear regime of our problem is similar to another curvature-driven problem -- that of the relaxation of a perturbed liquid-air interface \cite{tsaleza2013,benzaquen2014approach} -- in that they flow to attracting set of shapes at long times. We show this explicitly by simulating five different asymmetric initial-conditions each with a straight portion attached to a semi-circular portion as in fig.~\ref{fig:asymm_data}$(e)$. The total length of the filament for all the initial conditions is held fixed while the diameter, $b$  of the semi-circular section is varied. We quantify the asymmetry in terms of the difference in curvature on either side of the midpoint:
$
\phi = \int_0^{{L_o}/2} |\k(L_o/2-s) - \k(s)| ds
$.
$\phi$ takes a value of 0 for a completely symmetric shape and positive values for different levels of asymmetry. We plot $\phi(t)$ in fig.~\ref{fig:asymm_data}. In all cases, the filament first rapidly becomes symmetric and then relaxes more slowly to a straight line. The straight section first curves in order to attain overall symmetry, thus showing that tension must play a role, as there is no bending force on the straight section. The time-scale to reach the symmetric shape increases monotonically with $b$. (see fig.~\ref{fig:asymm_appx} of appendix to see the movement of $T(s)$ along the filament to regions of zero bending force.) 
 
We make an energy argument to show that the energy of the symmetric state is a minimum with respect to asymmetric perturbations. Let us assume a smooth symmetric profile for curvature, $\hk(s)=\hk(-s)$.  We perturb the curvature, $\hk(s)$, while maintaining the boundary conditions,  with a function  $-\epsilon(s)$ in $s \in [-L_o/2,0]$ and with $\epsilon(-s)$ in $s \in [0,L_o/2]$ . The total elastic energy becomes:
\begin{align*}
\ee &= \frac{B}{2} \bigg[ \int_{-L_o/2}^0 (\hk(s) - \epsilon(s))^2 ds + \int_{0}^{L_o/2} (\hk(s) + \epsilon(-s))^2 ds \bigg]\\
&= \frac{B}{2} \bigg[ \int_{-L_o/2}^{L_o/2} \hk^2(s) ds + \int_{0}^{L_o/2}(\epsilon^2(-s)+\epsilon^2(s))ds \bigg]\\
& \ge \frac{B}{2} \int_{-L_o/2}^{L_o/2} \hk^2(s) ds
\end{align*}
It is thus evident that for a given curvature, a symmetric shape has minimum energy. This shows that asymmetric perturbations about an arbitrarily large-deformation state will be relaxed to a symmetric state. To emphasize the fact that it is a symmetric stress or curvature, rather than a symmetric shape, that is attained, we show an example in fig.~\ref{fig:asymm_data}$(c,d)$ where $\r(s)$ is not symmetric but $\kappa(s)$ is. These initial conditions also relax along a stress-symmetric sequence of shapes. The fast relaxation of the asymmetric stress state to a symmetric one is to be expected because the effective length of the portion of the filament under tension is shorter at early times, and changes with time as the tension gets distributed everywhere in the filament.

\section{Conclusion}
\sgp{We have found that the dynamics of the relaxation process may be collapsed over the whole range of the dynamics by a single time scale arising out of balancing viscous drag and bending force.  This time-scale has been previously recognized [\cite{quennouz2015transport},\cite{2006PhFl...18i1701Y},\cite{2008PhFl...20e1703C}] from the governing equations, but we show here that the observed   relaxation time in the nonlinear regime is much shorter than this time scale, presumably because the tension in the filament and nonlinear bending lead to a shape-dependent prefactor to the time scale.  A qualitative explanation for this is currently lacking, but we can show that the relaxation as measured by the end-to-end distance, for instance, is governed by large tension gradients (see eq.~\ref{eq:lt}, and data in fig.~\ref{fig:forces}).}


 \sgp{We are developing this experimental  setting  to  measure  interfacial  viscosity  by  measuring  the  relaxation  of a  filament  with  known  elastic  properties  and  geometry.
 As an example, we are currently studying the time-dependence due to water absorption of the interfacial viscosity of glycerol.} Alternatively, once the drag at an interface is calibrated for a known elastic rod, this setting can be used to infer the bending modulus of a filament. The preference for symmetric shape of the filament over asymmetric shapes would make the experiment robust to small variations in initial conditions. 

\section{Appendix}
\subsection{Expression for $L(t)$}
The evolution of vectors at the boundary of the filament can be written as:
\begin{align}
\bar{\mu}\partial_t \r|_0 &= - \r_{ssss}|_0 + [\r_s\partial_sT(s)]|_0\\
\bar{\mu}\partial_t \r|_L &= - \r_{ssss}|_L + [\r_s\partial_sT(s)]|_L
\end{align}
For symmetric initial conditions that we use in experiments, we can write:
\begin{align*}
\r_{ssss}|_0 &= \r_{ssss}|_L\\
\partial_sT(s)|_0 &= -\partial_sT(s)|_L
\end{align*}
Now using these the vector connecting the ends evolves as,
\begin{align}
\bar{\mu}\partial_t (\r|_0 - \r|_L) &=  \partial_sT(s,t)|_0[\r_s|_0+\r_s|_L]\\
\bar{\mu}\partial_t \bar{L}(t) {\x} &= 2\partial_sT(s,t)|_0 \cos(\theta(t)) {\x}
\label{eq:lt}
\end{align}
where $\theta(t)$ is the angle between the tangent vector at $s=0$ and the horizontal. This shows that the end-to-end distance depends only on the gradient of tension at the boundary and bending is eliminated from the equation.
\subsection{Independence of height of glycerol}
\begin{figure*}
\centering
\includegraphics{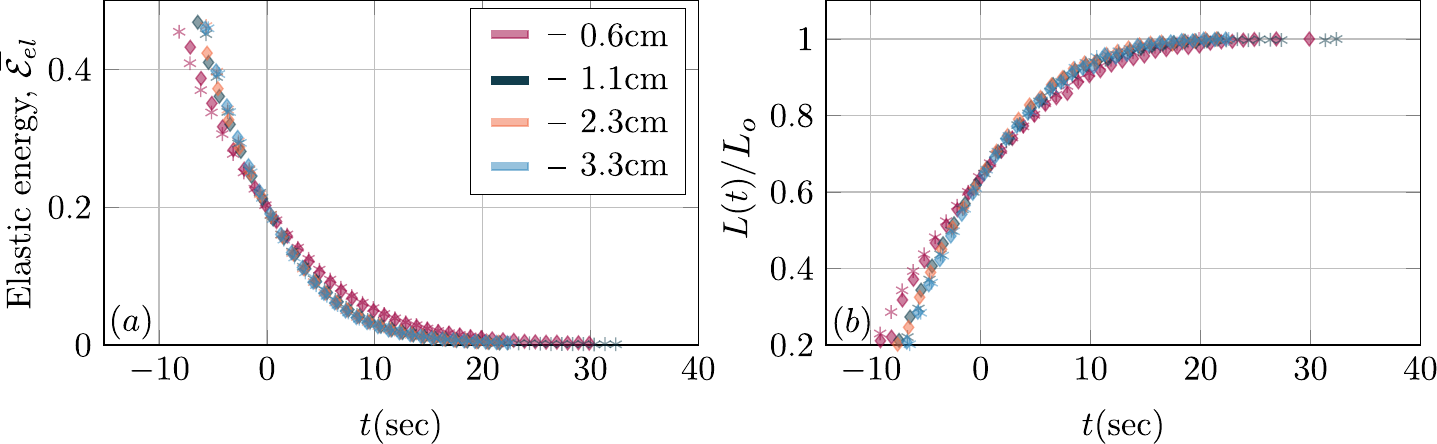}
\caption{$(a)$ Energy, $(b)$ end-to-end distance for different heights of glycerol. We do not see any significant effect except at the shallowest height of $0.6 cm$.}
\label{fig:height_glycerol}
\end{figure*}

One might be concerned that the relaxation dynamics would be affected by the depth of the glycerol in the container. In our experiments, the filament is placed at the glycerol-air interface. We report here the minimum height of glycerol required in the beaker for the filament relaxation dynamics to be not affected by the volume of glycerol in the container. Fig.~\ref{fig:height_glycerol} shows the relaxation for four different height: $h=0.6 cm, 1.1 cm, 2.3 cm, 3.3 cm$. We see that for height greater than $0.6cm$, the dynamics remain the same. Thus in all our experiments, a height greater than $3.3cm$ is used.

\subsection{Effect of hydrodynamic interaction}
\begin{figure*}
\centering
\includegraphics{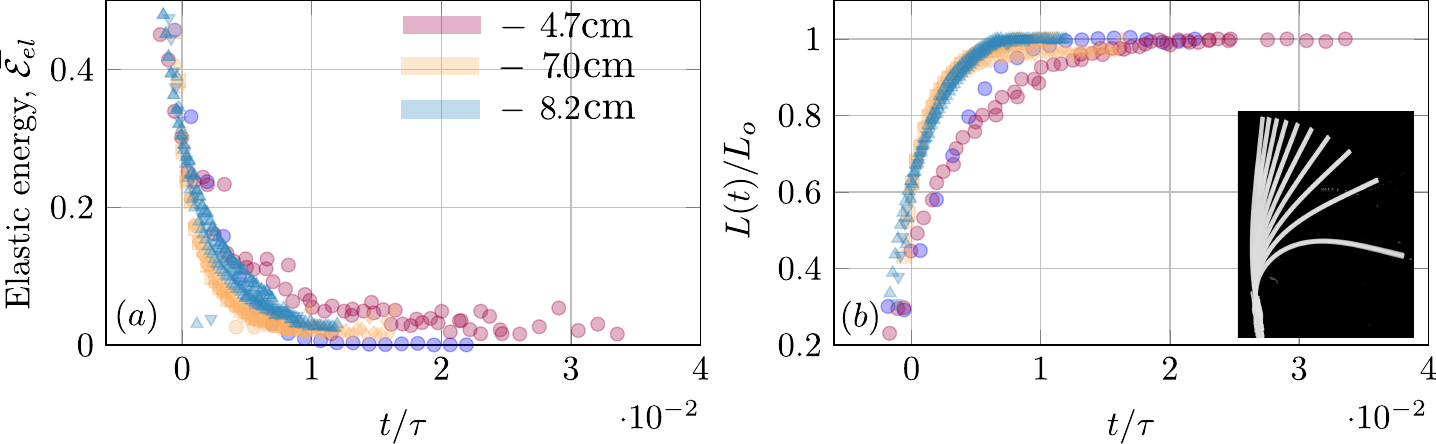}
\caption{$\nee$ and $L(t)/L_o$ vs $t/\tau$ for a filament with one end clamped and the other end deformed. $L(t)$ is obtained by mirroring the image about the normal at the hinged end. The effective length calculated by combining the mirrored image and the normal relaxation is represented by $L_o$. Blue circles correspond to relaxation of the symmetric initial configuration described earlier.}
\label{fig:cantilever}
\end{figure*}

The model used to simulate the filament neglects the hydrodynamic interaction between different points along the filament as mentioned in the main text. But these interactions are present in the experiments and thus to see if the relaxation is radically modified by these interactions, we perform the following experiment. We clamp one end of the filament and deform the other end and let it relax from this configuration. The relaxation is mirrored about the normal at the fixed end and the effective length of this combined filament is $L_o$. Now $L(t)$ is calculated for this combined picture that consists both the mirrored part and the actual relaxation. Fig.~\ref{fig:cantilever} shows $\nee$ vs $t/\tau$ and $L(t)/L_o$ vs $t/\tau$ for three different lengths and we see that the collapse is spread. The blue rounds are that of the symmetric relaxation from earlier experiments. Though this does not conclusively quantify the effect of hydrodynamic interaction, we see that $\tau$ is not exorbitantly modified.

\begin{figure*}
\centering
\includegraphics[scale=0.35]{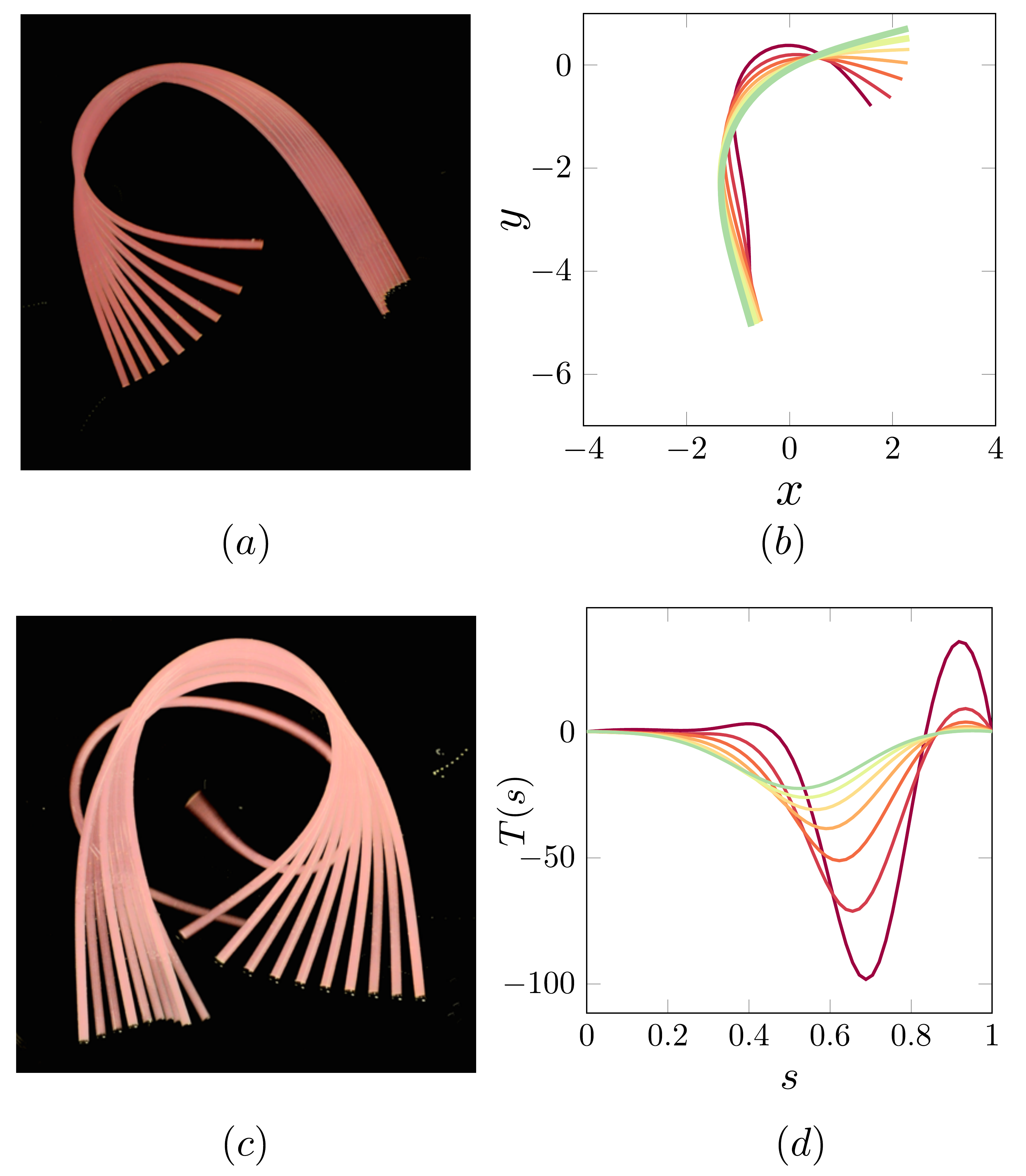}
\caption{$(a,c)$ Two arbitrary asymmetric initial conditions relaxing to symmetric states with $(b,d)$ showing similar behaviour from numerics. $(d)$ We see that the tension, $T(s)$ moves towards regions of zero bending force and becomes symmetric. Even after symmetrising, the tension is still of finite amplitude, indicating that the system is still non-linear. }
\label{fig:asymm_appx}
\end{figure*}

\begin{acknowledgements}
The authors would like to thank P.T. Brun, Arthur Evans and Joey Paulsen for important discussions and SGP wishes to thank the hospitality of people at UMass Amherst where part of this work was done. SGP was funded by APS-IUSSTF grant and JM by Raman-Charpak fellowship. We acknowledge funding from TCIS Hyderabad and NSF-DMR 120778 and 1506750 (NM).
\end{acknowledgements}

\bibliography{biblio}

\end{document}